\documentclass[10pt, conference, compsocconf]{IEEEtran}
\usepackage{cite}
\usepackage{float}
\usepackage{graphicx}
\usepackage{pifont}
\usepackage{bbding}
\usepackage{url}
\usepackage{hycolor}
\usepackage[cmex10]{amsmath}
\usepackage{amssymb}
\usepackage{amsthm}
\usepackage{stmaryrd}
\usepackage{bm}					
\usepackage{graphicx}		
\usepackage{tikz}				
\usepackage{booktabs,ctable,multirow}
\usepackage{mdframed}
\usepackage{stfloats}
\usepackage[bookmarks,
bookmarksopen,
bookmarksnumbered,
colorlinks,
hypertexnames=false,		
urlcolor = {blue}
]{hyperref}							
\usepackage{xcolor}			

\DeclareFontFamily{U}{msb}{}
\DeclareFontShape{U}{msb}{m}{n}{ <5> <6> <7> <8> <9> gen * msbm
  <10> <10.95> <12> <14.4> <17.28> <20.74> <24.88> msbm10}{} 
\DeclareSymbolFont{AMSb}{U}{msb}{m}{n}
\DeclareMathSymbol{\E}{\mathalpha}{AMSb}{"45}
\DeclareMathSymbol{\I}{\mathalpha}{AMSb}{"49}
\DeclareMathSymbol{\N}{\mathalpha}{AMSb}{"4E}
\DeclareMathSymbol{\R}{\mathalpha}{AMSb}{"52}
\DeclareMathSymbol{\Z}{\mathalpha}{AMSb}{"5A}

\newcommand{\atk}[2][k]{\ensuremath{#2}^{({#1})}}
\newcommand{\cerc}[1]{${\varbigcirc\mspace{-13mu} \text{\footnotesize #1}\;}$}

\newcommand{\sm}[1]{\mathcal{#1}}

\DeclareMathOperator{\argmax}{argmax}
\DeclareMathOperator{\prob}{Prob}

\definecolor{periwinkle}{RGB}{178,176,255}
\definecolor{oldpink}{RGB}{255,177,212}
\definecolor{violet}{RGB}{54,2,115}
\definecolor{marine}{RGB}{85,100,255}
\definecolor{pomegrenate}{RGB}{255,18,94}
\definecolor{cyan}{RGB}{0,255,255}
\definecolor{vert}{RGB}{11,194,85}
\definecolor{noir}{RGB}{0,0,0}

\newcommand{\bn}{\bm{n}}

\newcommand{\bw}{\bm{w}}
\newcommand{\bz}{\bm{z}}

\newcommand{\bI}{\bm{I}}

\newcommand{\bP}{\bm{P}}

\newcommand{\bU}{\bm{U}}

\newcommand{\bze}{\bm{\zeta}}

\newcommand{\sZ}{\bm{\mathsf{Z}}}

\begin{document}
\title{Decoding Epileptogenesis in a Reduced State Space} 

\author{ Fran\c{c}ois G. Meyer$^1$, Alexander M. Benison$^2$, Zachariah
  Smith$^2$, and Daniel S. Barth$^2$\\
  $^1$Electrical Engineering \& Applied Mathematics, $^2$Psychology \& Neuroscience\\
  University of Colorado at Boulder, Boulder, CO 80309 }
\maketitle
\begin{abstract}
  We describe here the recent results of a multidisciplinary effort to design a biomarker that can
  actively and continuously decode the progressive changes in neuronal organization leading to
  epilepsy, a process known as epileptogenesis. Using an animal model of acquired epilepsy, we
  chronically record hippocampal evoked potentials elicited by an auditory stimulus. Using a set of
  reduced coordinates, our algorithm can identify universal smooth low-dimensional configurations of
  the auditory evoked potentials that correspond to distinct stages of epileptogenesis. We use a
  hidden Markov model to learn the dynamics of the evoked potential, as it evolves along these
  smooth low-dimensional subsets. We provide experimental evidence that the biomarker is able to
  exploit subtle changes in the evoked potential to reliably decode the stage of epileptogenesis and
  predict whether an animal will eventually recover from the injury, or develop spontaneous
  seizures.
\end{abstract}
\section{Introduction}
\subsection{The Challenge: Decoding Epileptogenesis}
Epilepsy is a neurological disease that is characterized by the occurrence of several unprovoked
seizures. Despite various causes of epilepsy and varying degrees of disease severity in the human
population, hippocampal sclerosis is the most consistent neuropathological feature of temporal lobe
epilepsy \cite{pitkanen11}.  Animal models have been developed to study the neuronal changes
underlying the clinical manifestations of epilepsy (chronic-spontaneous seizures). One popular model
relies on controlled administration of a convulsant drug (e.g., pilocarpine) to induce
status-epilepticus, a life-threatening condition in humans. This condition is followed by a latent
seizure-free period of weeks to months, where progressive neuronal damage and network reorganization
eventually lead to the development of spontaneous seizures. Most of our understanding of the
progression of epilepsy, or {\em epileptogenesis}, is derived from such animal models. It is
therefore critical to define a {\em biomarker} to monitor epileptogenesis. An accurate biomarker
would be invaluable for the design of novel anti-epileptogenic drugs, and could eventually be
translated into a diagnostic tool for humans.  
%
\subsection{From Passive Recording to Active Sensing}
Current efforts toward the development of a reliable and predictive biomarker of epileptogenesis
fall in three main classes: molecular and cellular biomarkers \cite{lukasiuk14}, imaging biomarkers
\cite{shultz14}, and electrophysiological biomarkers \cite{staba14}. The present work focuses on the
last class of biomarkers that rely on recordings of the electrical activity associated with neuronal
firing. The development of electrophysiological biomarkers has focused on the analysis of both
epileptiform spikes \cite{huneau13}, and high frequency oscillations \cite{engel12}.  Interictal
spikes are sharp electrical impulses. Their morphology has been shown to be correlated with the
progression of epileptogenesis \cite{huneau13}. Recent experiments combined with computational
models \cite{huneau13} suggest that epileptogenesis systematically modifies the morphology of the
spikes in a predictable manner.  High frequency oscillations detected in local field potentials are
created by synchronously bursting pyramidal cells, and have been observed in healthy patients when
the frequency of the ripples is lower than 250 Hz. Conversely, frequencies in the range of 250Hz to
800Hz are considered to be pathological ripples and are reliable biomarkers for the epileptogenic
zone \cite{delaprida15}. While recordings of spontaneous neuronal spiking can be indicative of
neuronal excitability, and therefore correlate with the propensity for seizures, such methods cannot
actively probe the hippocampal circuit in living animals during epileptogenesis. Consequently, one
could argue that the {\em passive} electrophysiological recordings may not provide enough
information to observe early changes in neuronal excitability associated with epileptogenesis.

The present work addresses this limitation and proposes for the first time a
``computational biomarker'' that relies on actively probing the excitability of the hippocampus
using an auditory stimulus. We advocate an {\em active} approach whereby we probe the excitability of the
hippocampus, a brain region known to be epileptogenic. Because the hippocampus also receives several
sensory inputs (through the entorhinal cortex), we propose to record in the hippocampus the evoked
potential elicited by an auditory stimulus. Since epilepsy does not modify the primary auditory
cortex, any alterations in the evoked potential should be indicative of neuronal changes in the
hippocampus underlying epileptogenesis. To quantify the property of this biomarker, we use a
pilocarpine animal model of temporal lobe epilepsy, and chronically record hippocampal auditory
evoked potentials during epileptogenesis. 

We design a decoding algorithm to demonstrate that changes in the morphology of the hippocampal
auditory evoked potential have universal predictive value and can be used to accurately quantify the
progression of epilepsy. The authors are not aware of any work that uses machine learning methods to
construct an {\em active} biomarker of epileptogenesis (but see \cite{huneau13} for approaches based
on passive recordings). %
\begin{figure}[H]
  \centerline{
    \includegraphics[width=0.5\textwidth]{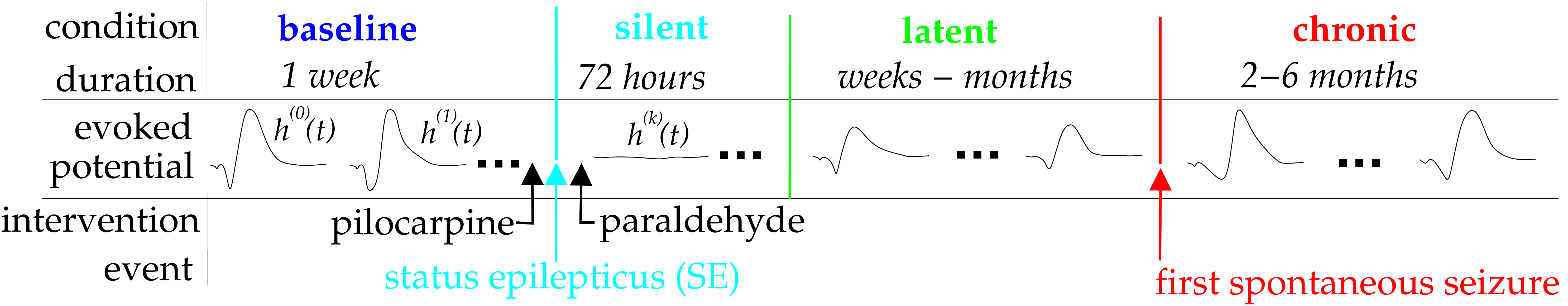}
  }
  \caption{Timeline and nomenclature of the different conditions.
    \label{timeline}}
\end{figure}
\vspace*{-1em}
\section{The Animal Model
\label{animal}}
All procedures were performed in accordance with the University of Colorado Institutional Animal
Care and Use Committee guidelines for the humane use of laboratory rats in biological research.
Twenty-four male Sprague-Dawley rats (200-250 gm) were implanted with a hippocampal wire electrode,
a ground screw, and a reference screw. The 24 rats were divided into two groups: a group of 17 rats
that received lithium-pilocarpine, and a control group of 7 rats. The control group was composed of
2 rats that received all drug injections associated with the lithium-pilocarpine model except for
pilocarpine, which was substituted with saline; and 5 rats that received no drugs.  After full
recovery from the electrode implantation (2 weeks) and at least one additional week of chronic
recording of baseline video/EEG, 17 rats were given an injection of lithium chloride followed by an
injection of pilocarpine hydrochloride 24 hours later (see Fig.~\ref{timeline}). After one hour of
status epilepticus, the animals were administered a dose of paraldehyde to terminate convulsions.
Throughout the experiment, every $\Delta t=30 \; \text{min}$, an auditory stimulus, composed of a
sequence of 120 square-wave clicks (0.1 ms duration, 2 sec ISI, 45dB SPL), was played in a
top-mounted speaker. The 300~ms hippocampal responses to each click were filtered and sampled at
10~kHz, and the average of the 120 responses was computed. In the remainder of the paper, we denote
by $\atk{h}$ the average evoked potential, measured at time $k\Delta t$. To further simplify the
exposition, $\atk{h}$ is simply referred to as the {\em evoked potential measured at time $k$}.

Figure~\ref{timeline} provides a detailed timeline, along with the nomenclature of the different
periods associated with the progress and eventual onset of epilepsy. The period before the injection
of pilocarpine is called {\em baseline}. Conversely, the period following the first spontaneous
seizure is called {\em chronic}. We further define the {\em silent} period to be the 72 hour period
of recovery immediately following the termination of status epilepticus, and the {\em latent} period
to be the remaining period leading to the eventual onset of the first spontaneous seizure.
\section{Overview of the Decoding Framework}
\begin{figure}[H]
  \centerline{
    \includegraphics[width=0.5\textwidth]{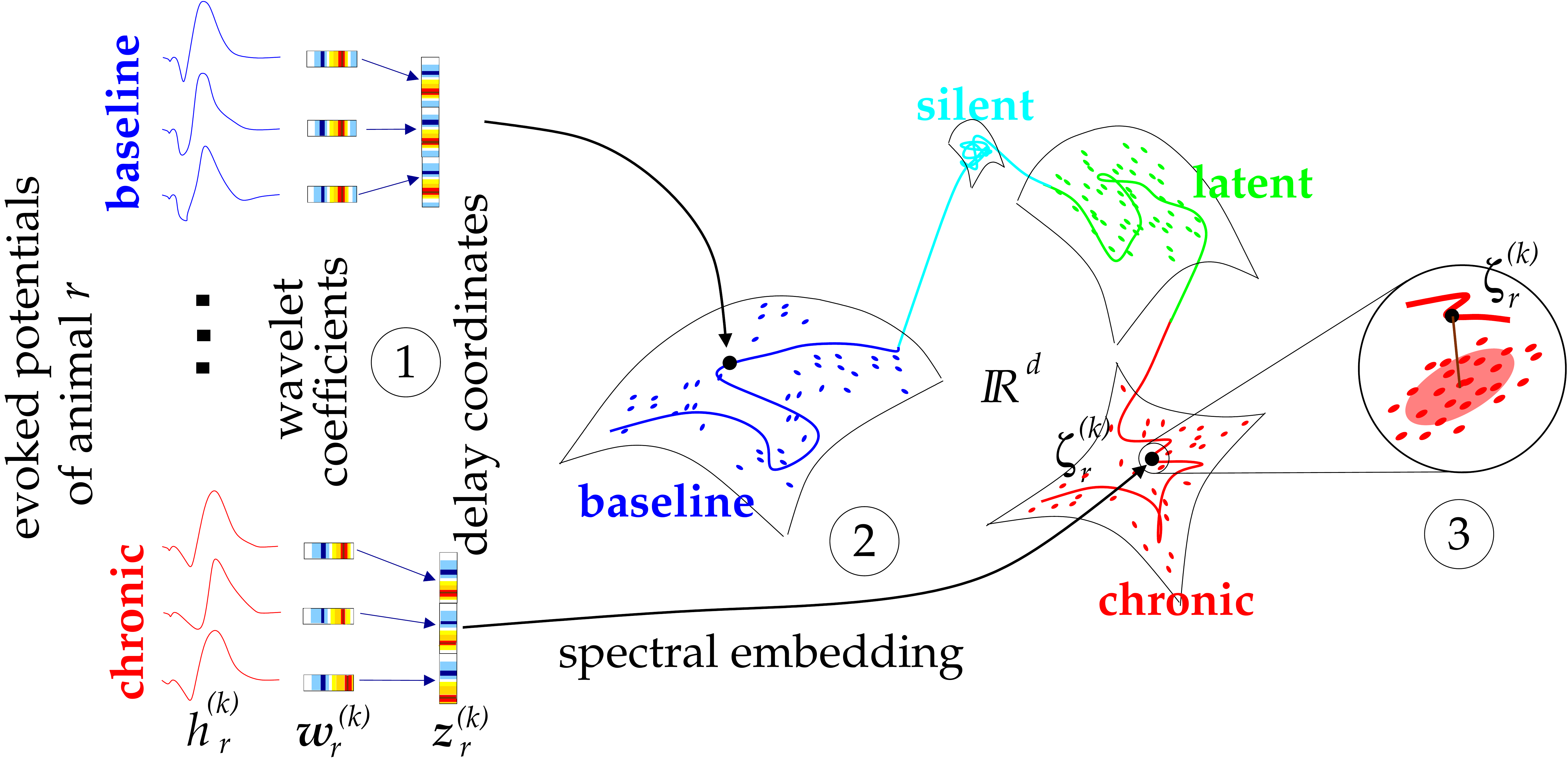}
  }
  \caption{Overview of the decoding algorithm. \cerc{1}: a vector of wavelet coefficients,
    $\atk{\bw}_r$, is computed from the evoked potential $\atk{h}_r$. A vector of time-delay wavelets
    coordinates, $\atk{\bz}_r$, is formed by concatenating $\tau$ consecutive
    $\atk{\bw}_r$. \cerc{2}: spectral embedding maps $\atk{\bw}_r$ to $\atk{\bze}_r$. \cerc{3}: the
    distance between $\atk{\bze}_r$ and the low-dimensional structure formed by each condition is
    computed. 
    \label{framework}}
\end{figure}

We give here a brief overview of the decoding approach. Given an animal $r=1,\ldots,24$, we consider
the sequence of evoked potentials $\atk[0]{h}_r, \atk[1]{h}_r, \ldots$.  The first stage involves
the construction of a denoised representation of $\atk{h}_r$ (see \cerc{1} in
Fig.~\ref{framework}). We decompose $\atk{h}_r(t)$ using a discrete stationary wavelet transform and
retain a vector of $s$ (carefully chosen) 
\vspace*{-0.7em}
\noindent wavelet and scaling coefficients, $\atk{\bw}_r$, (see
\cerc{1} in Fig.~\ref{framework}).  The second stage involves characterizing the association between
the condition of the disease (baseline, silent, latent, or chronic) and the vector of wavelet
coefficients $\atk{\bw}_r$.  We tackle this question by lifting $\atk{\bw}_r$ into
$\R^{\tau \times s}$ using time-delay embedding: we concatenate the consecutive vectors
$\atk{\bw}_r, \ldots, \atk[k+\tau-1]{\bw}_r$ to form a $\tau \times s$ vector, $\atk{\bz}_r$, of
{\em time-delay wavelet coordinates} (see \cerc{1} in Fig.~\ref{framework}). Low-dimensional
structures, which uniquely characterize the stage of epileptogenesis, emerge in the high-dimensional
space. We use spectral embedding to parameterize these low-dimensional structures, and map
$\atk{\bz}_r$ to $\atk{\bze}_r$ (see \cerc{2} in Fig.~\ref{framework}). The first decoding stage
involves geometrically computing the likelihood that a given vector $\atk{\bze}_r$ corresponds to
one of the four conditions. To this end, we quantify the distance of $\atk{\bze}_r$ to the
low-dimensional cluster formed by each condition (see \cerc{3} in Fig.~\ref{framework}).  In the
final decoding stage, we use a hidden Markov model to capture the intrinsic dynamics of
epileptogenesis.

To alleviate the notation, and unless we explicitly compare or combine several animals, we dispense
with the subscript $r$, denoting the dependency on rat $r$, when we discuss the analysis of the
evoked potentials for a fixed animal.
\section{Denoising the Input Data
 \label{denoise}} 
To compensate for the alteration of the tissue impedance around the electrodes, the evoked
potentials $\atk{h}_r, k = 0,1,\ldots$ were normalized for each animal. For a given animal $r$, all
evoked potentials for that animal were rescaled, in such a way that the average energy computed
during the baseline condition, $\langle h_r^2\rangle$, was one.  In order to use the noisy evoked
potentials to predict the state of epileptogenesis, we extract a denoised representation of
$\atk{h}$. We use a discrete stationary wavelet transform (CDF 9-7) to compute a redundant
representation of $\atk{h}$. We used a ``leave-one-animal-out'' cross-validation to determine the
time intervals and the scales of the wavelet coefficients that resulted in the
\begin{figure}[H]
\centerline{\footnotesize \hfill baseline \hfill silent \hfill latent \hfill chronic\hfill}
\vspace*{1em}
  \centerline{
    \includegraphics[width=0.130\textwidth]{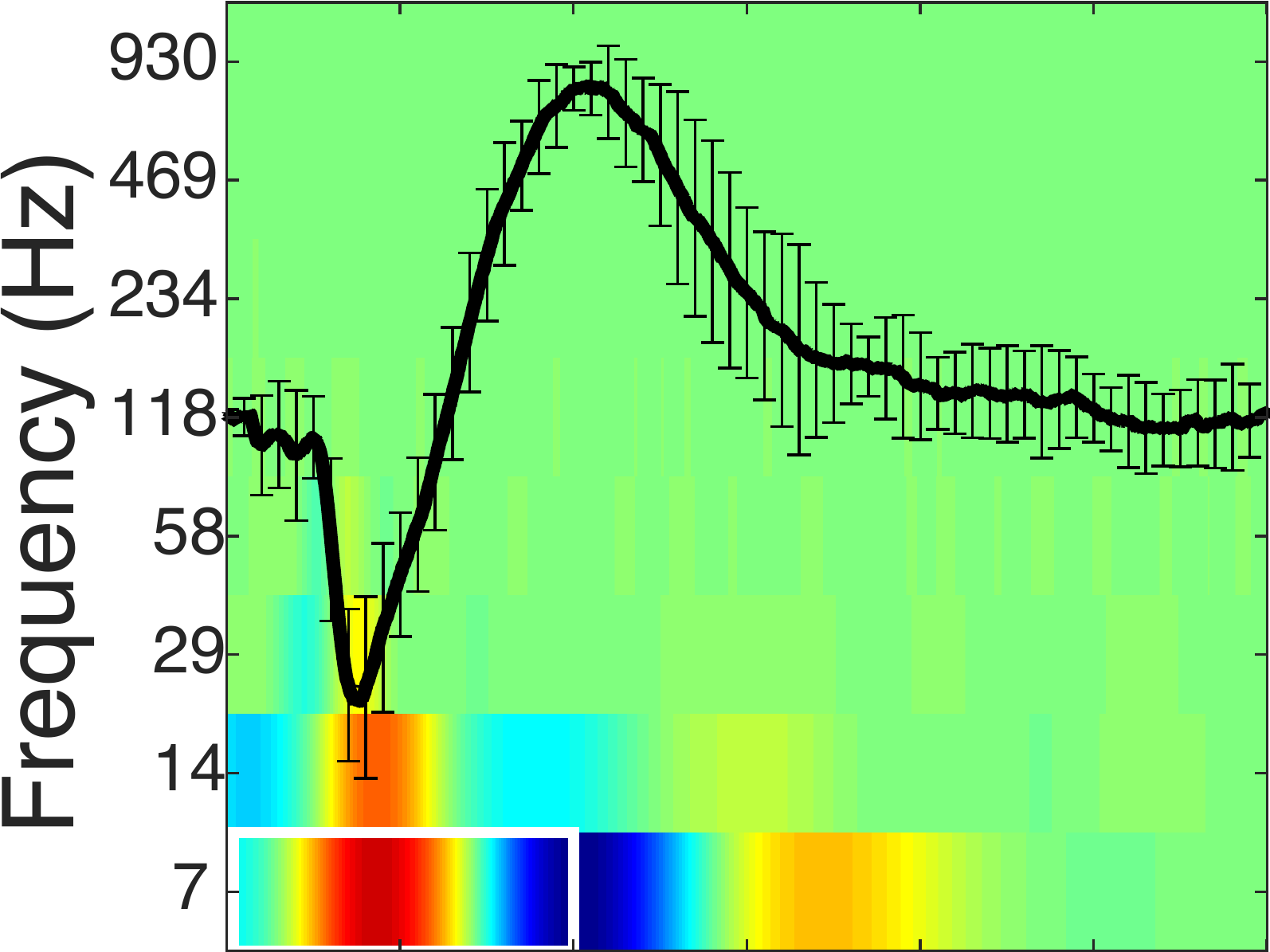}
    \hspace*{-0.2em}\includegraphics[width=0.1075\textwidth]{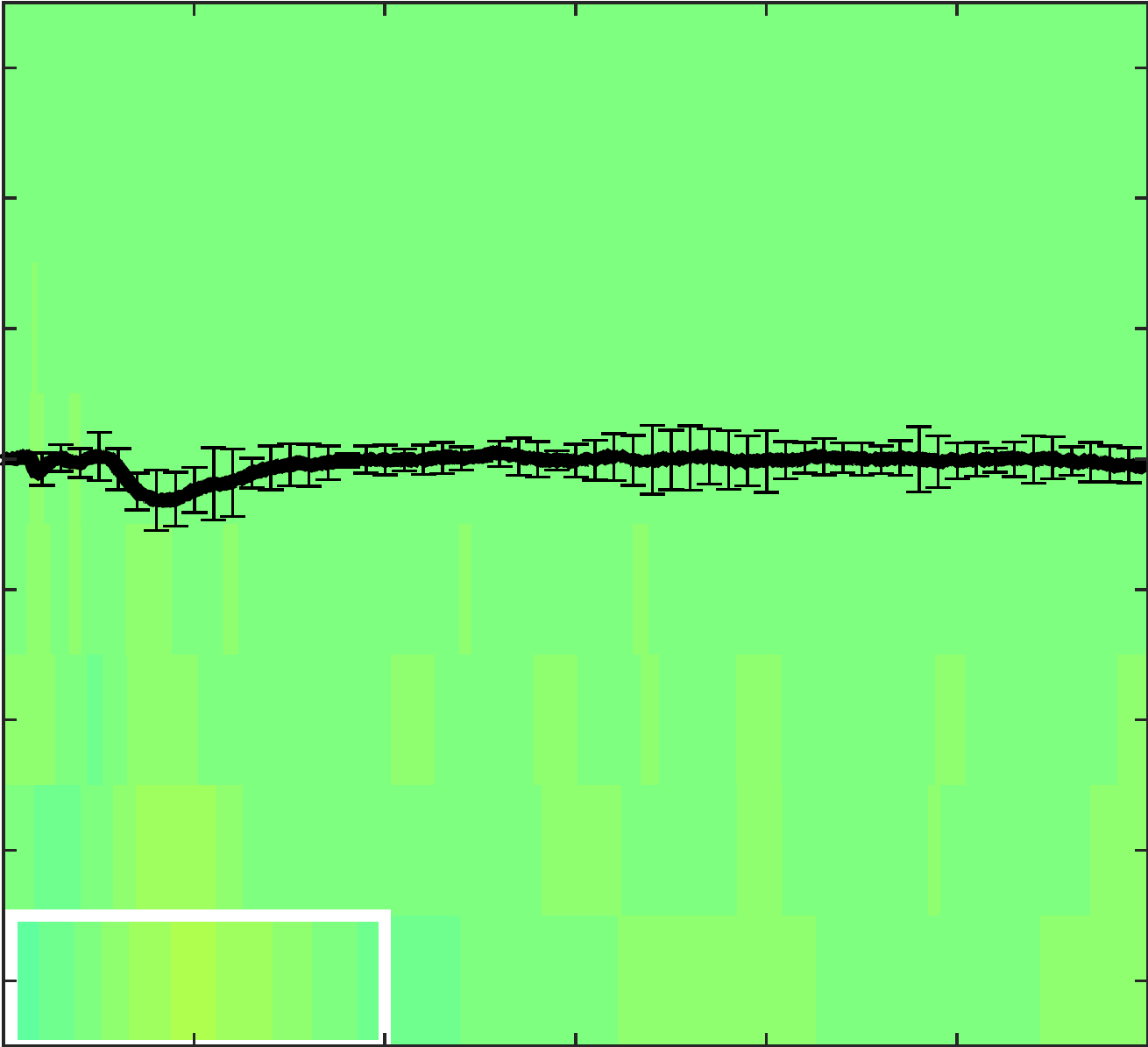}
    \hspace*{-0.2em}\includegraphics[width=0.1075\textwidth]{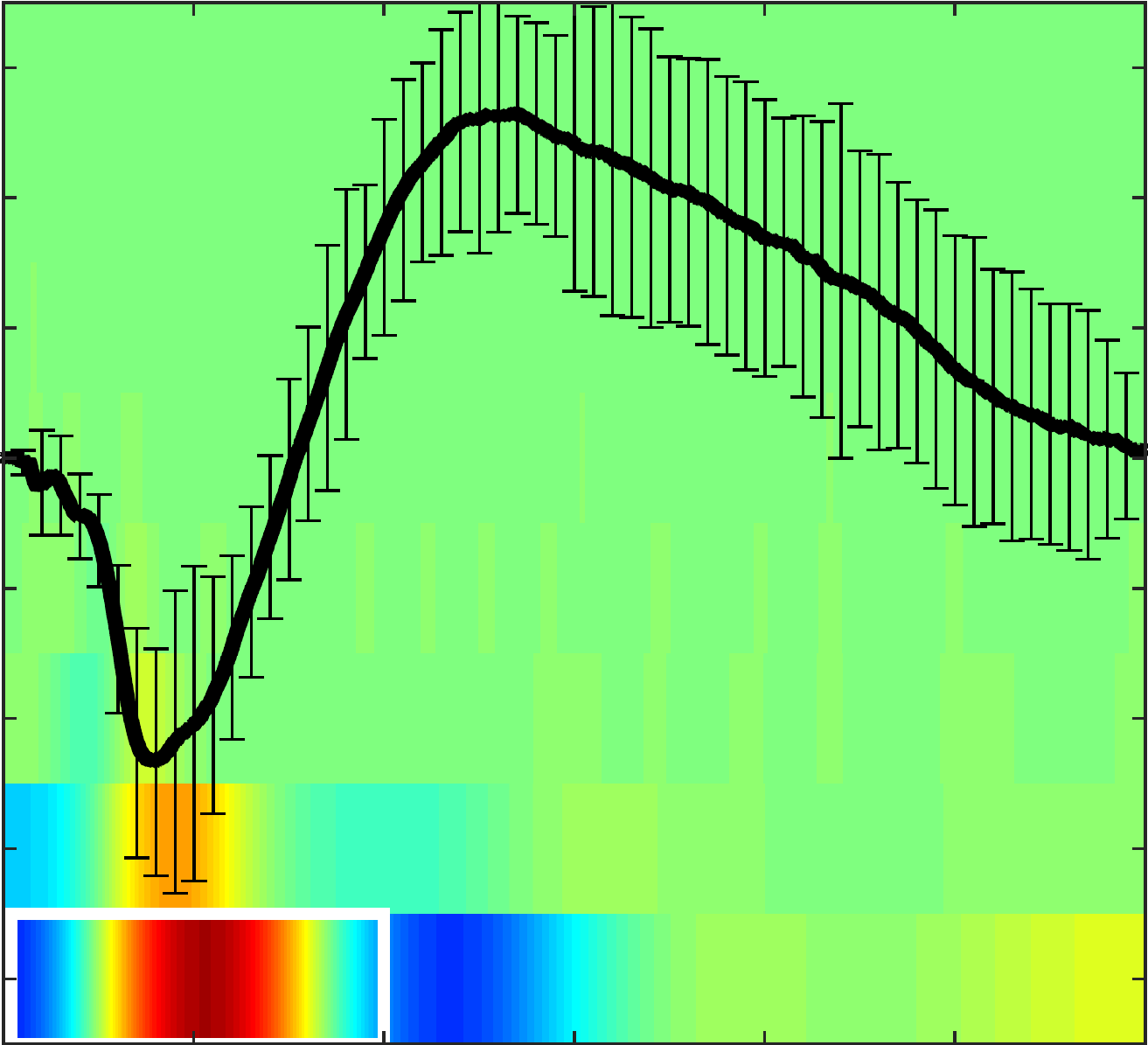}
    \hspace*{-0.2em}\includegraphics[width=0.135\textwidth]{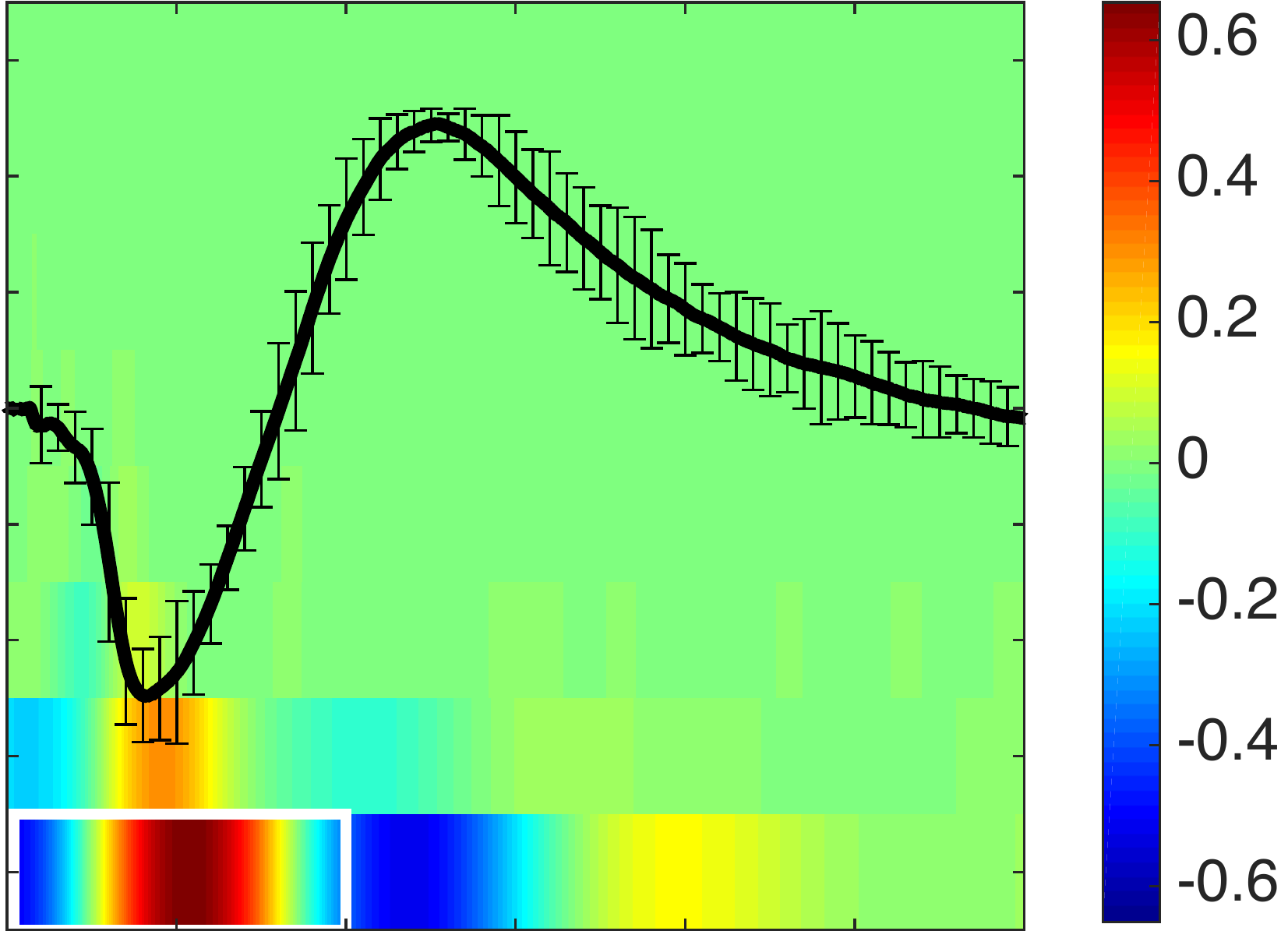}
  }
  \centerline{
    \includegraphics[width=0.130\textwidth]{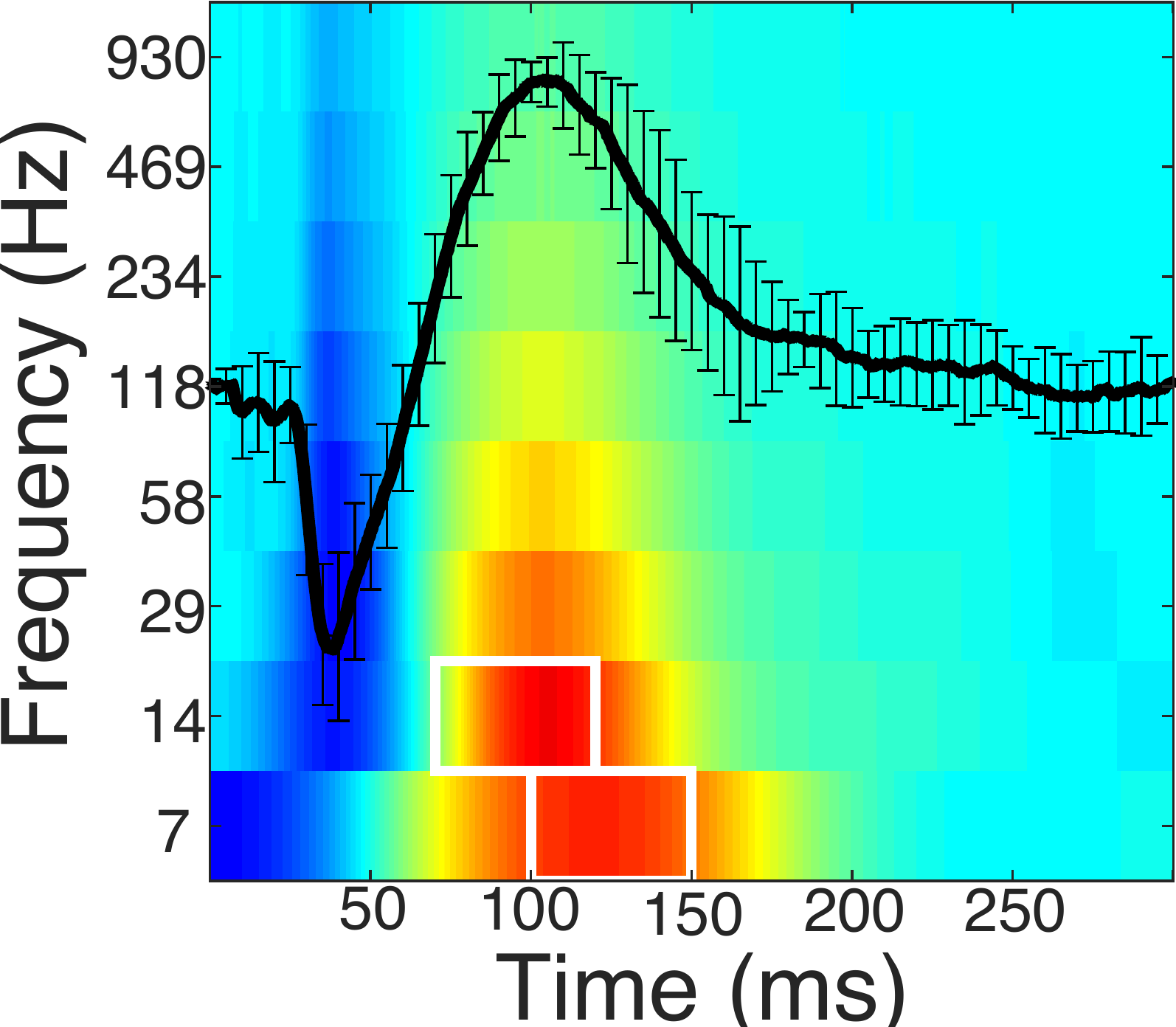}
    \hspace*{-0.2em}\includegraphics[width=0.1075\textwidth]{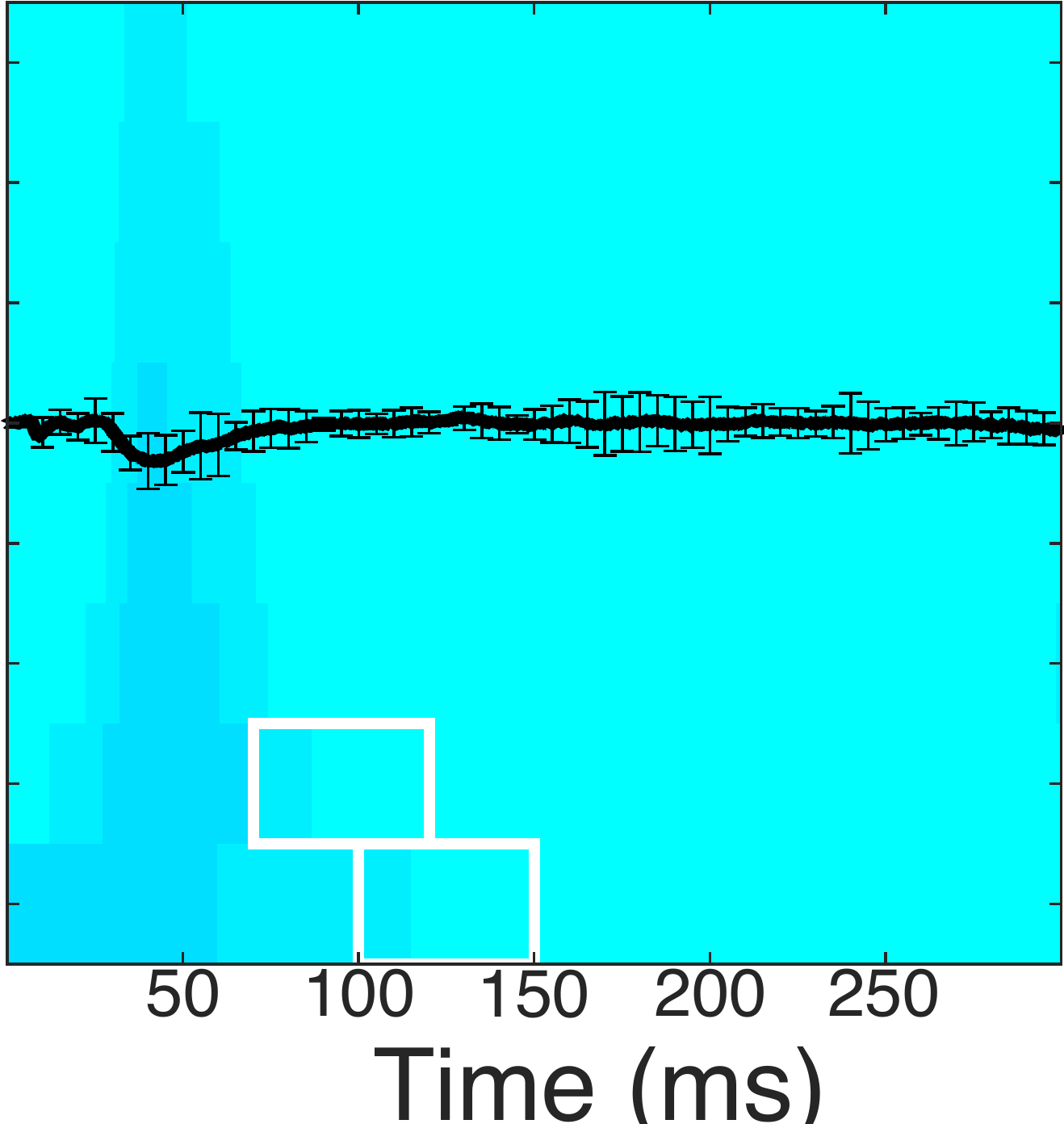}
    \hspace*{-0.2em}\includegraphics[width=0.1075\textwidth]{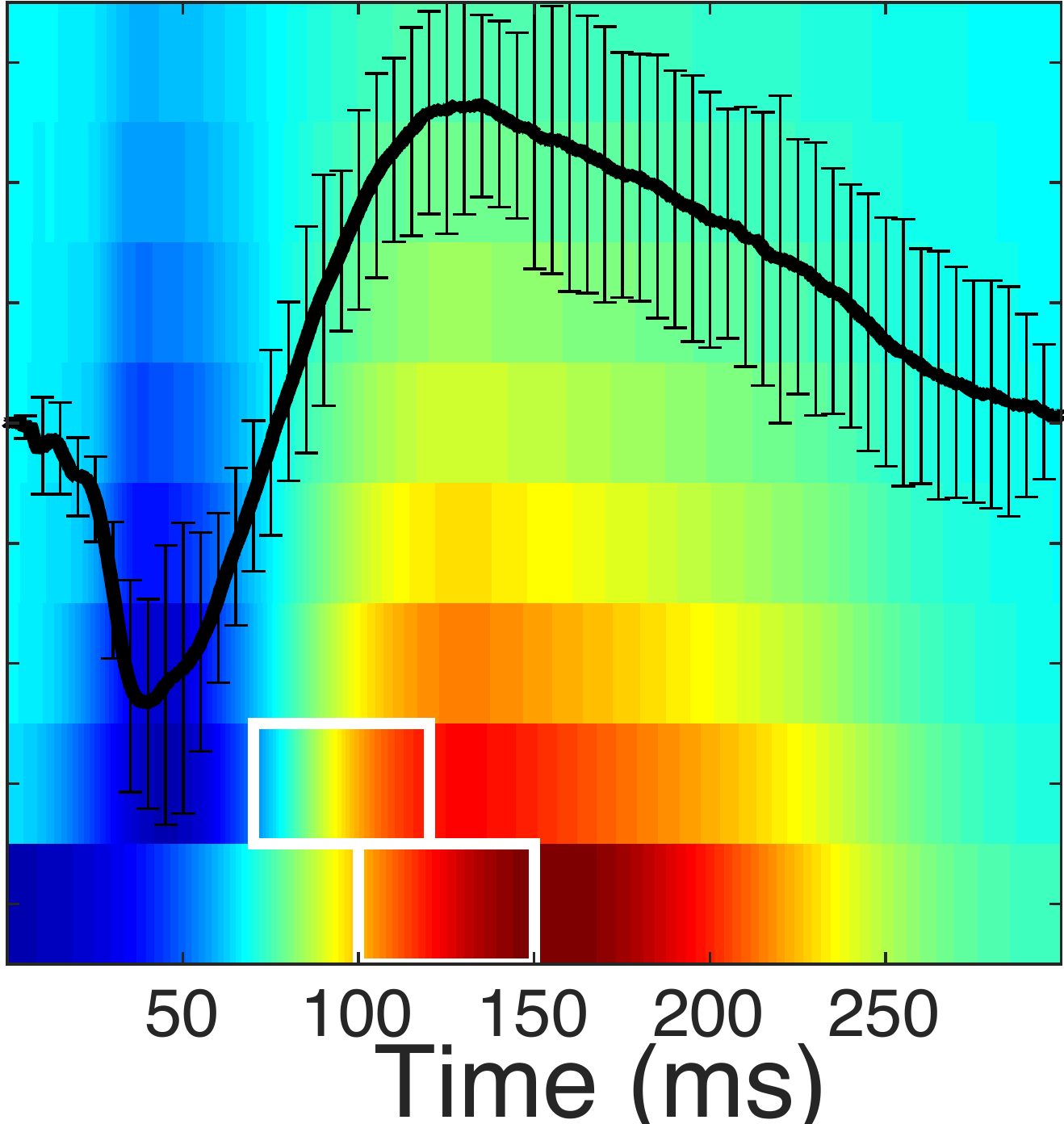}
    \hspace*{-0.2em}\includegraphics[width=0.135\textwidth]{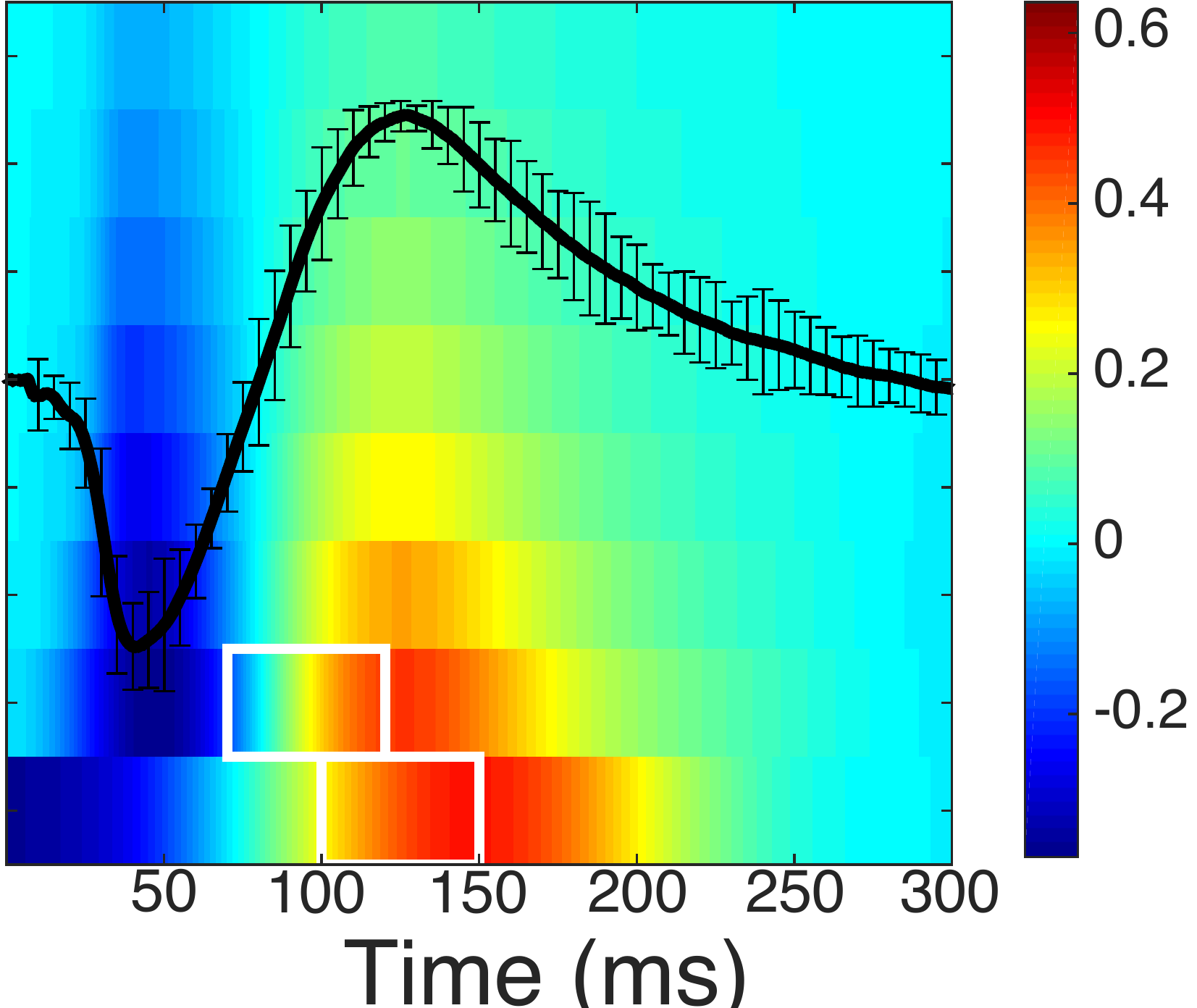}
  }
  \caption{Top: average wavelet coefficients with the average evoked potential (across all
    animals). Bottom: average approximation coefficients. Only the scales $j=3$ (top) to 10 (bottom)
    are displayed. White rectangles delineate the time-frequency blocks used to construct $\atk{\bw}$.
    \label{wavelet}}
\end{figure}
\vspace*{-.5em}
\noindent the most accurate prediction of the condition of the animal that was not part of the
training data. For most of the animals, the wavelet coefficients in the time-frequency region
$[0,100] \text{ ms} \times [5,10] \text{ Hz}$, and the approximation coefficients from the
time-frequency regions $[70,120] \text{ ms} \times [10,20] \text{ Hz}$ and
$[100,150] \text{ ms} \times [5,10] \text{ Hz}$ (see Fig.~\ref{wavelet}) resulted in either the
optimal, or very close to optimal, decoding performances. To simplify the decoding, we therefore
kept the same time-frequency regions for all animals. In summary, each evoked potential was
represented with a feature vector $\atk{\bw}$ of 2,000 entries composed of 1,000 wavelet
coefficients and 1,000 approximation coefficients. This representation is consistent with reports of
disruption in the $\theta$ rhythm (4-12 Hz) during the latent period preceding the onset of epilepsy
\cite{chauviere09}.
\section{Universal Configurations of Epileptogenesis
\label{embed}}
We now describe the construction of universal (stable across all animals) low-dimensional smooth
sets that are formed by all the hippocampal auditory evoked potentials collected during the same
stage of epileptogenesis. These sets are created by lifting the wavelet coefficients $\atk{\bw}$ of
each $\atk{h}$ into high-dimension.  This lifting effectively creates smooth low-dimensional
coherent structures that can then be parameterized with a drastically smaller number of coordinates.
\subsection{Time-Delay Embedding of the Wavelet Coordinates
\label{lifting}}
Given the time series $\left\{\atk{\bw}, k = 0,1,\ldots \right\}$ for a given animal, we analyze the
dynamics of this time series by considering  the time-delay wavelet coordinates formed by
concatenating $\tau$ consecutive vectors of wavelet coefficients,
\begin{equation}
  \atk{\bz}
  = 
  \begin{bmatrix}
    \atk[k]{\bw} & \atk[k+1]{\bw} & \cdots& \atk[k+\tau -1]{\bw}
  \end{bmatrix}.
  \label{wavelet-delay}
\end{equation}
We characterize the dynamical changes in the evoked potentials by studying the geometric structures
formed by the trajectory of $\atk{\bz}$ in $\R^{\tau \times s}$, as $k$ evolves. In practice, the
number of time-delay vectors, $\tau$, is determined using cross-validation. We confirmed that
$\tau=1$, to wit no time-delay
\begin{figure}[H]
  \centerline{
    \includegraphics[width = 0.2\textwidth]{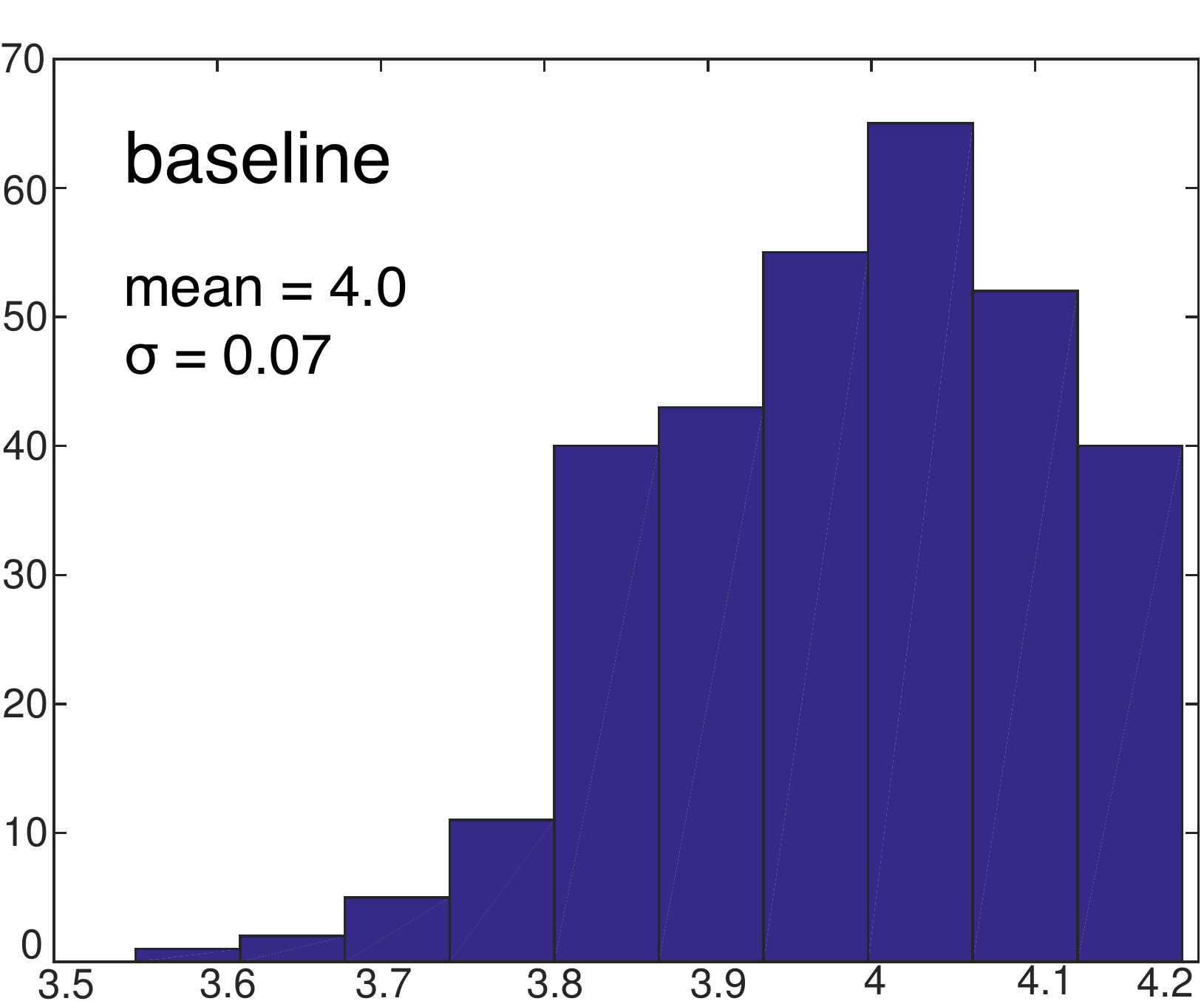}
    \includegraphics[width = 0.2\textwidth]{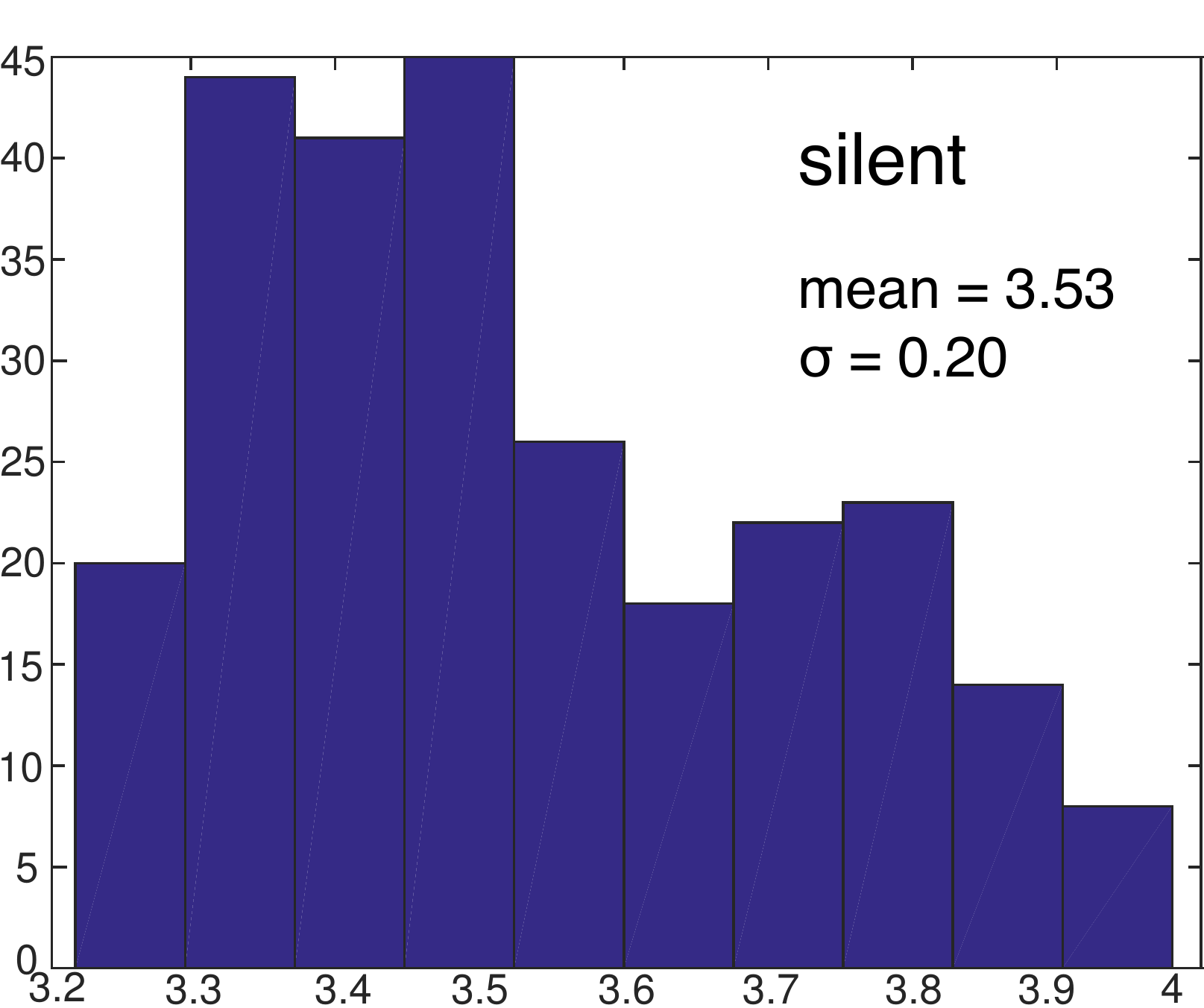}
  }
  \centerline{
    \includegraphics[width = 0.2\textwidth]{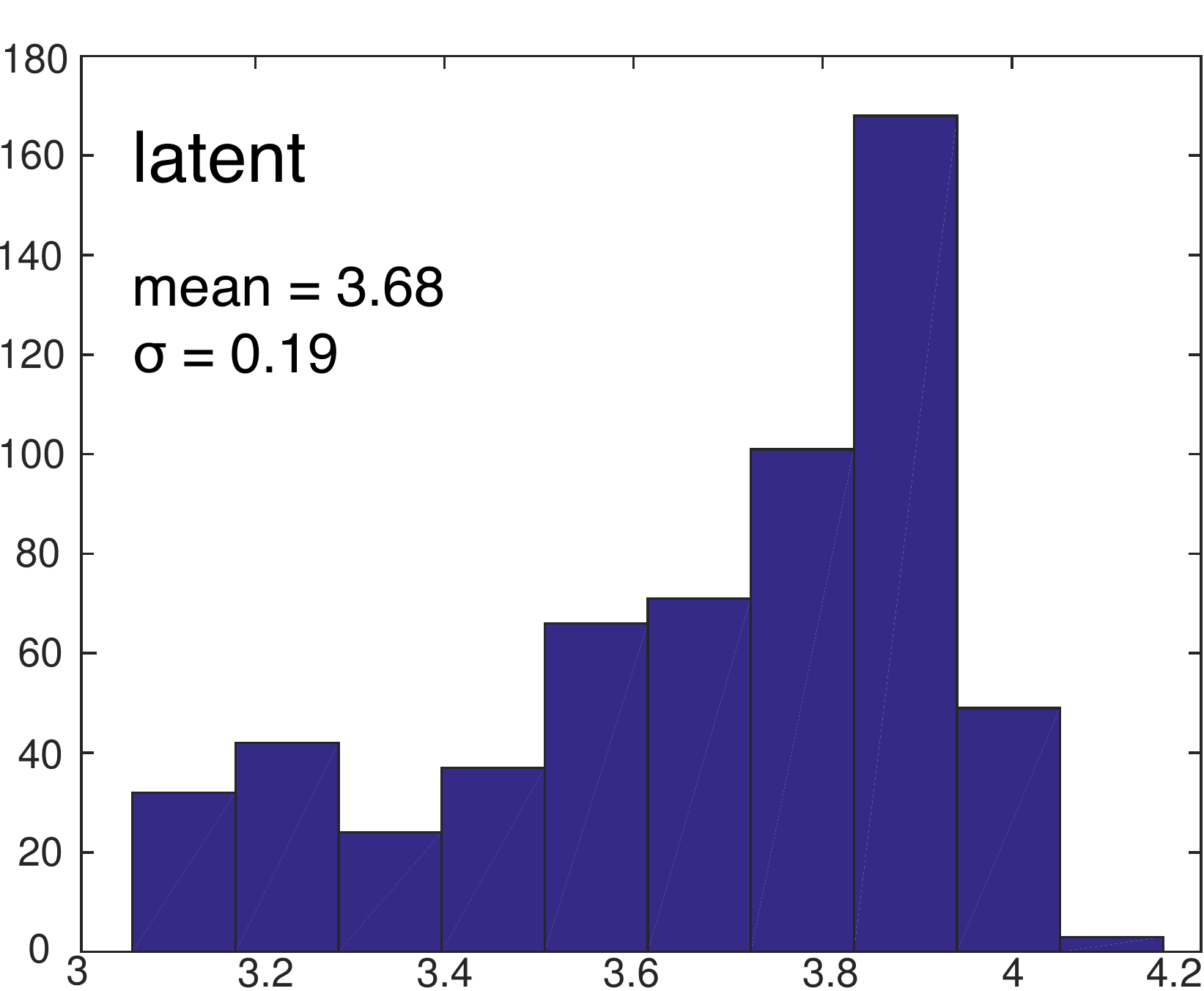}
    \includegraphics[width = 0.2\textwidth]{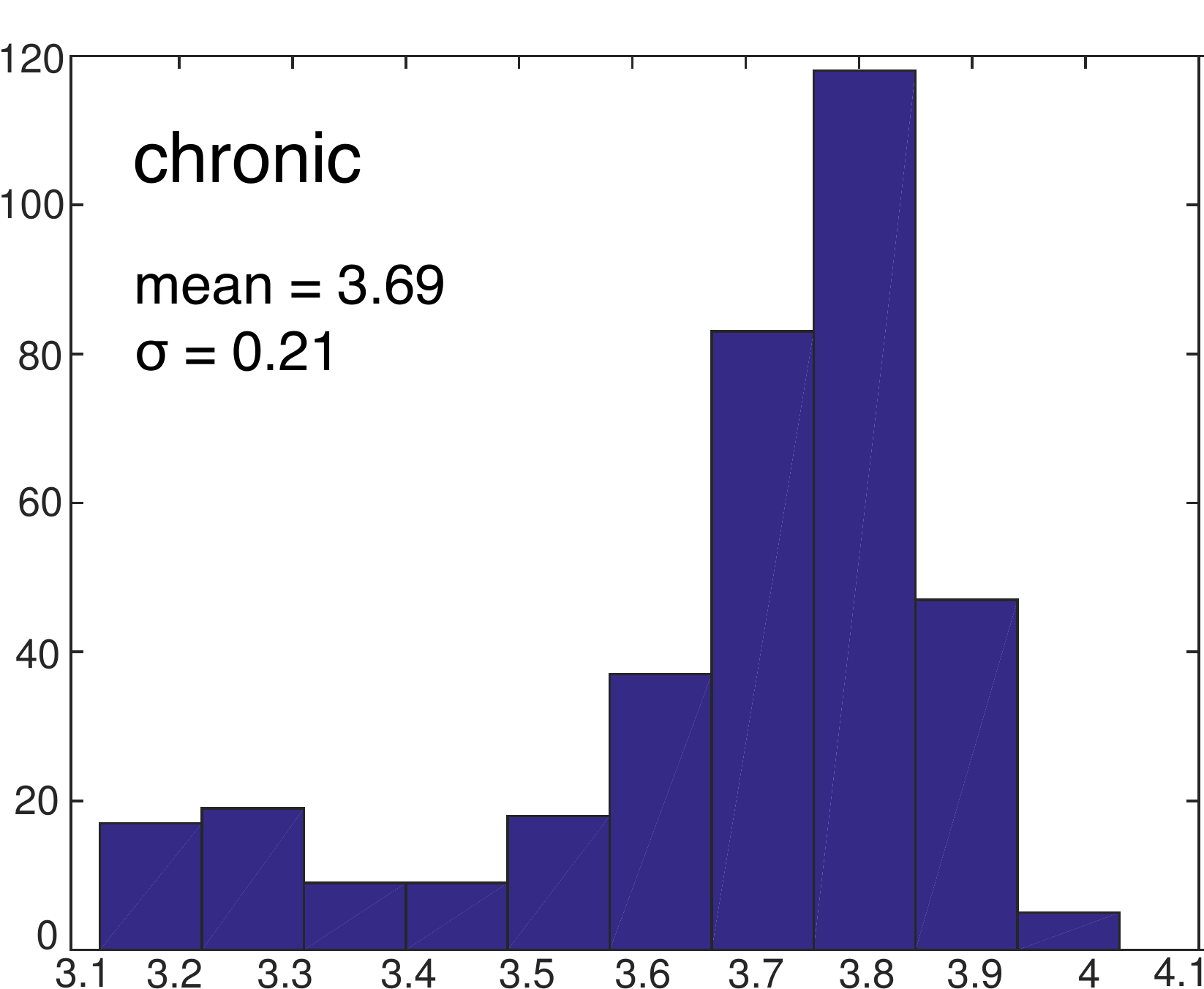}
  }
  \caption{Distribution of the local dimension of the state space according to the condition. Top:
    baseline and silent; bottom: latent and chronic.
    \label{local-dim}}
\end{figure}
\noindent  embedding, yielded poor prediction of the animal's condition. While
$\tau$ was optimized for each animal, the average of the optimal values was $\tau = 12$, which
corresponds to six hours.

To learn the geometry of the set formed by the different trajectories $\atk{\bz}_r, k =0,1\ldots$,
we consider the union -- over all the animals in the training set -- of the vectors $\atk{\bz}_r$,
and define the set 
\begin{equation}
  \sZ =   \mspace{-40mu} \bigcup_{\displaystyle r \in \text{\small training animals}}
  \mspace{-40mu}
  \left\{\atk{\bz}_r, k = 0,1,\ldots\right\}.
\end{equation}
\subsection{What is the local dimension of the state-space?
\label{localdim}}
Because we have only a limited number of training data, we need to drastically reduce the
dimensionality of the state space (on average $\tau \times s= 24,000$) in order to reliably decode
the condition of each animal. We used a kernel-based version of the correlation dimension
\cite{hein05} to compute the local dimension of the point-cloud formed by the state space $\sZ$. We
discovered that the points $\atk{\bz}_r$ were organized along low dimensional subsets that
correspond to well defined stages of epileptogenesis and exhibit different local dimensions. Figure
\ref{local-dim} displays histograms (as well as the mean, and standard deviation) of the estimates
of the local dimensions for the four different conditions.  These experiments suggest that we should
be able to represent the state-space associated with the dynamic of the disease using $d=5$
dimensions. However, our results (not shown) indicate that the traditional linear approach, PCA,
provides a very poor parameterization of the set $\sZ$.
\subsection{Nonlinear Parameterization of the State Space}
In order to identify the separate regions of $\R^{\tau \times s}$ that correspond to the four
conditions, we seek a smooth low-dimensional parameterization of $\sZ$. A nonlinear approach spectral embedding \cite{bengio06}, yields a low-dimensional parameterization of $\sZ$ that naturally
clusters the different conditions%
\begin{figure}[H]
\centerline{
  \includegraphics[width=0.5\textwidth]{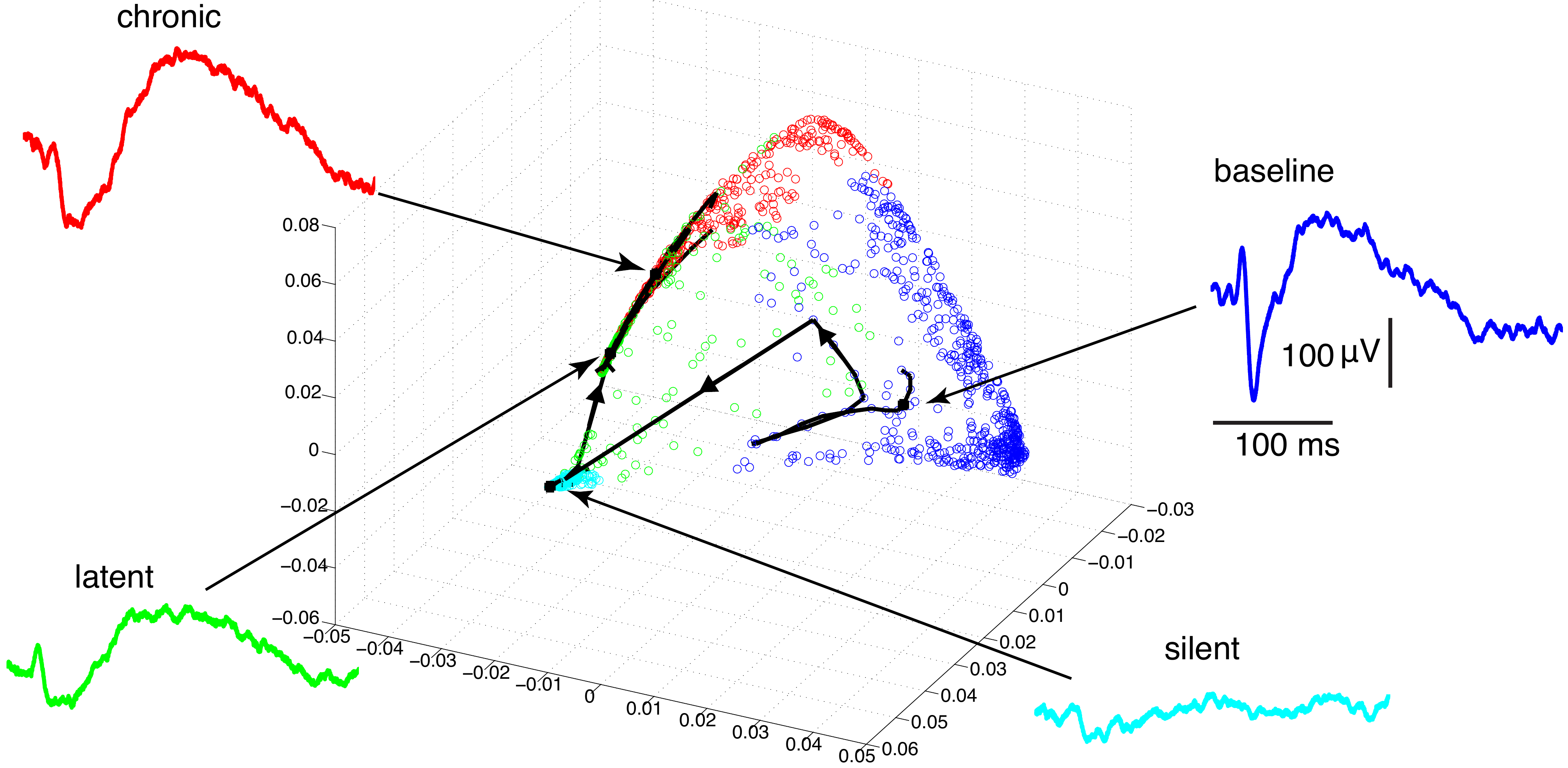}
}
\caption{The training set of evoked potentials, $\sZ$, displayed  using the reduced
  coordinates $\atk{\bze}$. Each  condition, baseline (blue), silent (cyan), latent (green), and
  chronic (red), forms a coherent sub-cloud. 
  \label{trajectory}}
\end{figure}
\noindent  into coherent disconnected smooth subsets.  Briefly, we define a
similarity matrix $\sm{K}$ that quantifies how any two evoked potentials $\atk[k]{h}$ and
$\atk[l]{h}$ extracted from the same or from different conditions, and from the same or from a
different animal, at the respective times $k$ and $l$ co-vary,
\begin{equation}
  \sm{K}(k,l) = \exp \left(-\|\atk[k]{\bz} - \atk[l]{\bz}\|^2/\sigma^2\right).
\end{equation}
The scaling constant $\sigma$ was chosen to be a multiple of the median distance
$\|\atk[k]{\bz} - \atk[l]{\bz}\|$, $k,l=0,1,\ldots$. We used the $d$ eigenvectors of $\sm{K}$ that
optimally separated the dataset $\sZ$ into four clusters  related to the four
conditions.  The $i^{\text{th}}$  eigenvector of $\sm{K}$ provided the $i^{\text{th}}$ {\em reduced
coordinate}, $\atk{\bze}(i)$, of $\atk{\bz}$,
\begin{equation}
\atk{\bze} =
\begin{bmatrix}
\atk{\bze}(1) & \cdots & \atk{\bze}(d)
\end{bmatrix}^T.
\end{equation}
As explained in section \ref{localdim}, we use $d=5$.  Figure~\ref{trajectory} displays the
nonlinear parameterization of $\sZ$. Each dot represents an evoked potential $\atk{h}_r$ measured at
a time $k$ from an animal $r$, and parameterized by the reduced coordinates $\atk{\bze}_r$.  The
color indicates the condition during which $\atk{h}_r$ was recorded. When collapsed across animals
and time of recordings, four distinct clusters can be visually discerned, which can be interpreted
in terms of the corresponding conditions. The evoked potentials in the silent condition are grouped
together around a low-dimensional structure. The latent condition displays the largest spread: some
evoked potentials are morphologically close to silent evoked potentials; while others are close to
chronic evoked potentials. These visual observations are  consistent with  the computation of the
local dimensionality described in section \ref{localdim}. To further help with the interpretation of
this nonlinear representation of $\sZ$, we trace the trajectory of $\atk{\bze}_r$, $k = 0,1,\ldots$
for a single rat (H37) over several weeks of recordings (see thick black line in
Fig.~\ref{trajectory}). The animal first spends several days in the baseline cluster (blue), then
briskly traverses the entire space to reach the silent cluster (cyan) after status epilepticus has
been induced. Eventually, as the animal recovers, it joins the latent condition (green), and
advances toward the chronic cluster (red).  Four representative evoked potentials are shown along
this trajectory, confirming changes in the morphology of $\atk{h}$ during epileptogenesis.
\subsection{Learning Epileptogenesis from the Reduced Coordinates}
The natural division of $\sZ$ into coherent subsets, which correspond to well defined stages of
epileptogenesis, suggests that a purely geometric algorithm could be used to quantify the
development of epilepsy. Indeed, given an unclassified evoked potential, $\atk{\bze}$, the distance
from $\atk{\bze}$ to each of the four clusters (baseline, silent, latent, and chronic) provides an
estimate of the likelihood of being in the corresponding condition at time $k$.

We denote by $\sZ^c\!\!,$ $c = 0, 1, 2, 3$, the four clusters formed by the evoked potentials in
$\sZ$ that were collected during the baseline, silent, latent, and chronic conditions,
respectively. We found that using a mixture of probabilistic principal components \cite{tipping99}
to represent each $\sZ^c$ resulted in a remarkably low dimensional parameterization of that
condition. Indeed, the mixture model is able to describe with a small number of components the
geometric structure created by each $\sZ^c$. Furthermore, unlike other models, the mixture model
does not require the stringent assumption that $\sZ^c$ be a smooth sub-manifold.

Formally, for each $c = 0, 1, 2, 3$, we use a different mixture of probabilistic principal
components to parameterize $\sZ^c$. To reduce the number of indices, we do not make the dependency on
$c$ explicit in the following discussion. For a given cluster $\sZ^c$, the mixture is composed of
$M$ local Gaussian densities, each describing the local principal directions with a $d \times q$
matrix, $U_i$, around a collection of centers $\overline{\bze}_i$, $i = 1,\ldots,M$.  The local
spread of the $\atk{\bze}_r$ around $\overline{\bze}_i$ is given by the covariance matrix
$U_iU_i^T + \varepsilon_i^2 \bI$. The mixture model associated with the cluster $\sZ^c$ can be
written as
\begin{equation}
  \atk{\bze}_r \sim 
  \sum_{i=1}^M  \pi_i {\cal N}
  \left(
    \bU_i  \bn_i + \overline{\bze}_i,\; \varepsilon_i^2  \bI
  \right),
\label{mixpca}
\end{equation}
where the $\bn_i, i = 1,\ldots,M$ are $q$-dimensional latent Gaussian variables, and the mixture
proportions $\pi_i$ are non negative and sum to one. The parameters of the model are estimated using
the EM algorithm \cite{tipping99}. For a given animal $r$, we compute the posterior probability that
the model (\ref{mixpca}) associated with condition $c$ would generate the reduced coordinates
$\atk{\bze}_r$,
\begin{equation}
\atk{\gamma}_r(c) = \prob(c | \atk{\bze}_r).
\label{firstproba}
\end{equation}
We use the posterior probability of class membership $\atk{\gamma}_r(c)$ as a raw measurement of the
progression of epileptogenesis. Because the instantaneous posterior probability $\atk{\gamma}_r(c)$
is%
\begin{figure*}[!t]  
  \centerline{
    \includegraphics[width=0.55\textwidth]{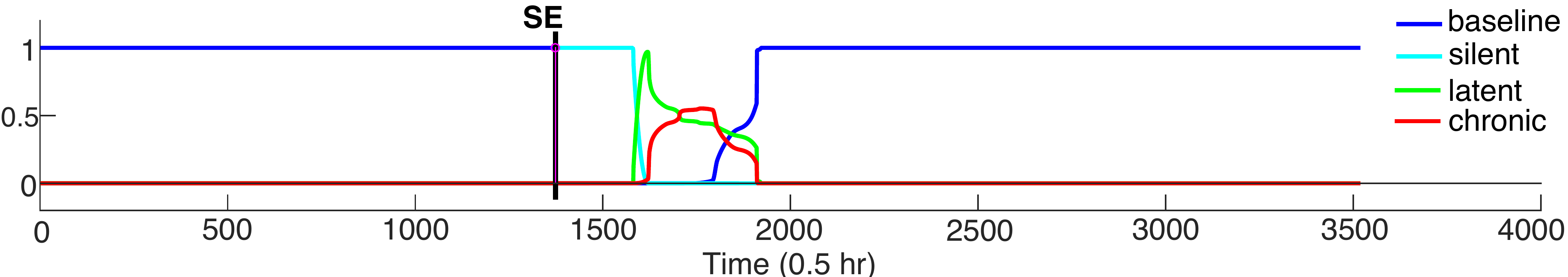}
  }
  \centerline{
    \includegraphics[width=0.6\textwidth]{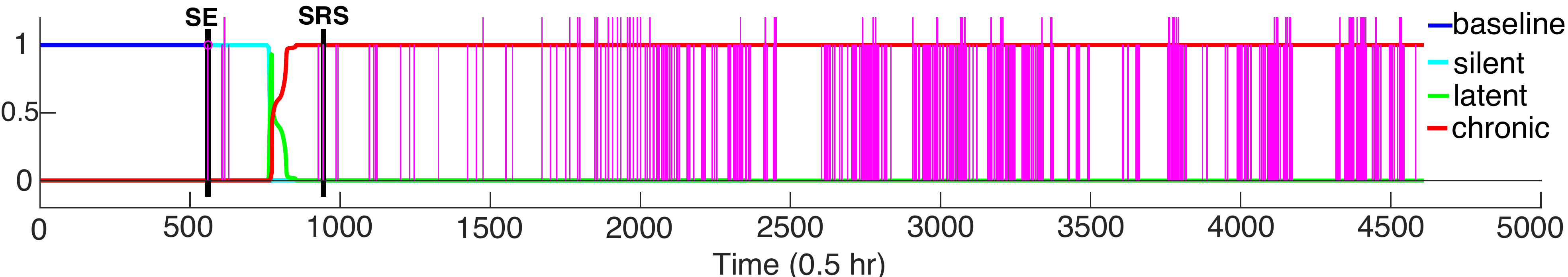}
  }
  \caption{Biomarker probabilities  ${\atk{\pi}}(c)$ of being in condition $c$ for a {\em non-epileptic} 
    (top) and an {\em epileptic} (bottom)  animal.   Seizure count (clipped at 1) is
    displayed in magenta. The black bar SE corresponds to {\em Status Epilepticus}; the second
    bar SRS corresponds to the first Spontaneous Recurring Seizure.
    \label{H40-PI2-decode}}
\end{figure*}
\noindent  oblivious to the past trajectory $\{\atk[k-1]{\bze}_r, \atk[k-2]{\bze}_r, \ldots \}$, it can be
noisy. To address this limitation, and enforce the temporal coherence that defines the progression of the disease, we introduce a Markovian
description of the dynamics of the changes in the evoked potential $\atk{h}$ as a function of time.
\section{The Dynamics of Epileptogenesis
  \label{dynamics}}
We use a hidden Markov Model to capture both the dynamics of epileptogenesis, and the resulting
changes in the shape of $\atk{h}_r$ triggered by this hidden condition. We define $\atk{x}_r$ to be
a discrete random variable that encodes the state of animal $r$ at time $k$,
$\atk{x}_r\in \{0, 1, 2, 3\}$, where the states 0, 1, 2, and 3 encode the baseline, silent, latent,
and chronic conditions, respectively. We further assume that $\atk{x}_r$ is a Markov chain with
probability transition matrix $P_{i,j} = \prob(\atk[k+1]{x}_r = i| \atk{x}_r = j)$. We define
$\atk{y}_r$ to be a discrete random variable, which takes its values in $\{0, 1, 2, 3\}$, and
encodes the most likely condition of animal $r$ at time $k$, given the measurement $\atk{h}_r$,
\begin{equation}
\atk{y}_r = \argmax\displaylimits_{c = 0, 1, 2, 3} \; \atk{\gamma}_r(c),
\end{equation}
where $\atk{\gamma}_r(c)$ is the posterior probability that condition $c$ generates the reduced
coordinates $\atk{\bze}_r$, and is given by (\ref{firstproba}). We denote by
$Q_{i,j} = \prob(\atk{y}_r = i| \atk{x}_r = j)$ the measurement probability distribution matrix. The
probability transition matrix $P$ and the measurement probability matrix $Q$ were estimated from the
training data. The trained hidden Markov model was then used in a ``reverse mode'' \cite{cappe09} to
estimate the posterior probability of animal $r$ to be in condition $c$,
\begin{equation}
  \atk{\pi}_r(c) = \prob(\atk{x}_r = c \;|\; \atk{y}_r), \quad 
c = 0, 1, 2, 3. 
\label{posterior}
\end{equation}
Because the hidden Markov model was trained on the epileptic animals, we regularized the 
probability transition matrix $\bP$ to allow a transition, which was never observed in the training
data, from chronic to baseline with a very small probability.

Figure~\ref{H40-PI2-decode} displays the four probabilities $\atk{\pi}$ of being in baseline,
silent, latent, and chronic state for a {\em non-epileptic} animal (top), and an {\em epileptic}
animal (bottom).  We plot the seizure count (clipped at 1) in magenta for the epileptic animal. The
biomarker clearly identifies the baseline period (blue), and is able to detect changes in the evoked
potential $\atk{h}$ that are indicative of neuronal alterations leading to epilepsy, before the
first spontaneous recurring seizure (SRS), and shifts from latent (green) to chronic (red) before
SRS.  The probability of being in the chronic state, $\atk{\pi}(3)$, remains at one after SRS,
indicating that the animal is irreversibly in the chronic state, as confirmed by the uninterrupted
sequence of seizures.
\section{Results}
\begin{figure*}[!t]  
  \centerline{
    \includegraphics[width=.3\textwidth]{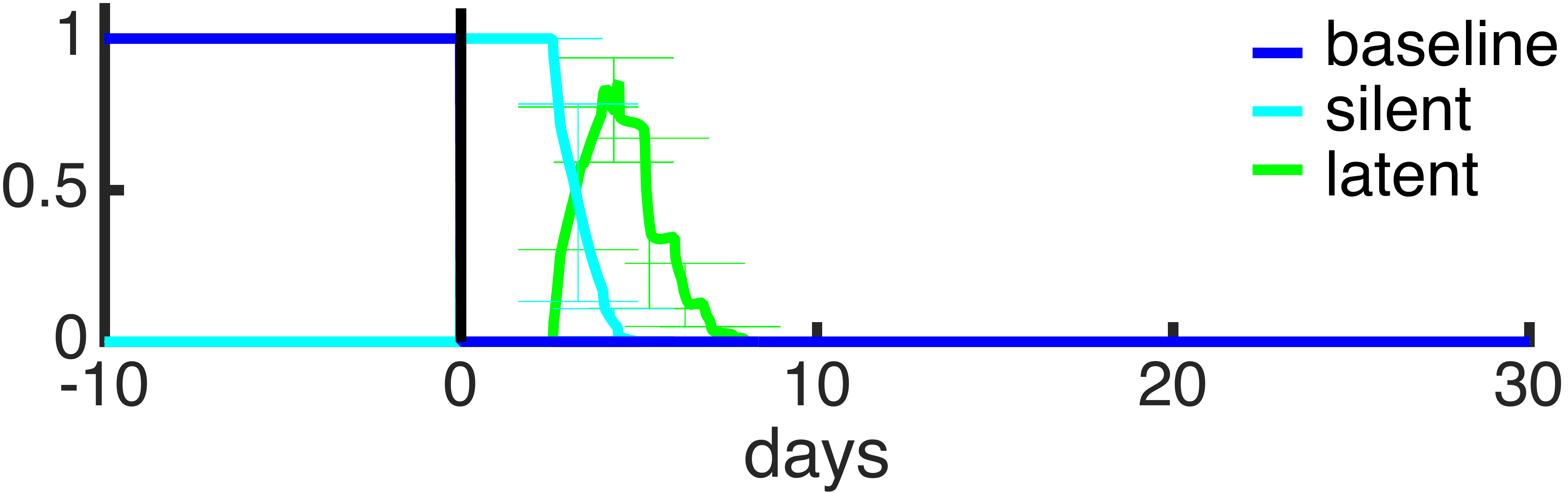}
    \includegraphics[width=.3\textwidth]{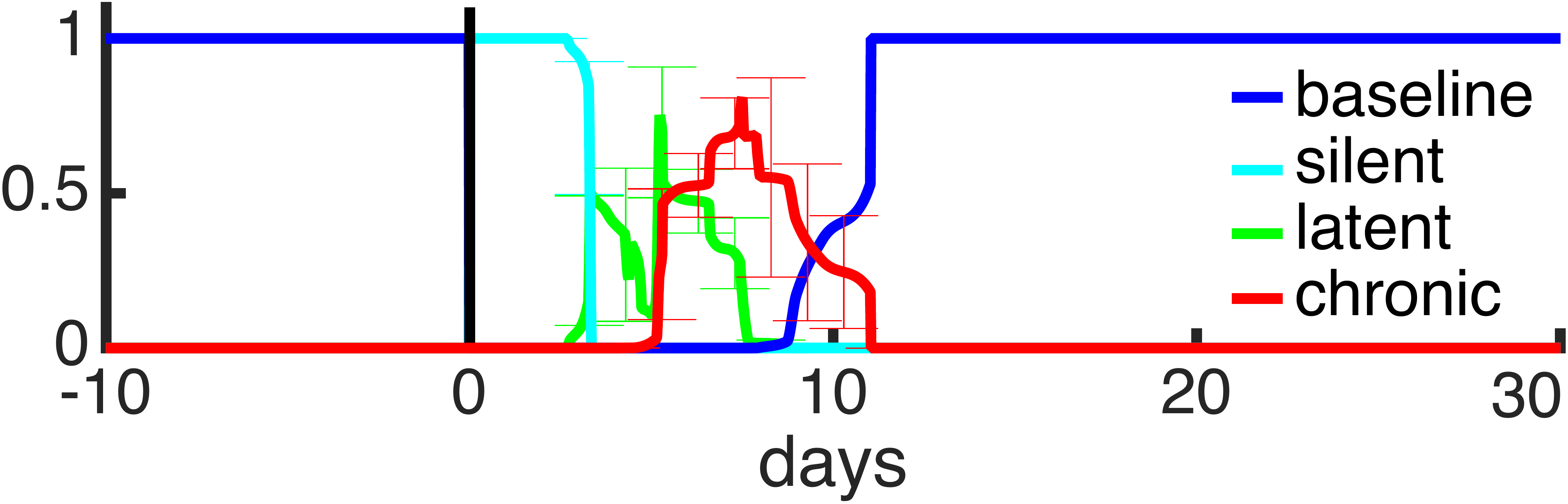}
  }
  \caption{Median probability $\overline{\atk{\pi}}(c)$ of being in condition $c$ for the epileptic animals (left),
    and the non epileptic animals (right). Day 0 (black bar) corresponds to status epilepticus across all animals.
    \label{after_se}}
\end{figure*}

We evaluated our approach using a ``leave-one-animal-out'' cross-validation: the algorithm was
trained on 11 epileptic rats, and evaluated with one animal that was not part of the training
data. For each test animal $r$, the decoding algorithm computed from $\atk{h}_r$ the posterior
probability $\atk{\pi}_r(c)$, given by (\ref{posterior}). Figure~\ref{after_se} displays the
temporal profile of the median probability $\overline{\atk{\pi}}(c)$ computed over all the epileptic
animals (left, N=12), and the non epileptic animals (right, N=5). The computation of the median
probability $\overline{\atk{\pi}}(c)$ required care: because each condition varied in length across
the different animals, we could not use the beginning of each recording as the origin, and bluntly
average across the animals.

\begin{figure*}[!t]  
\centerline{
    \includegraphics[width=0.6\textwidth]{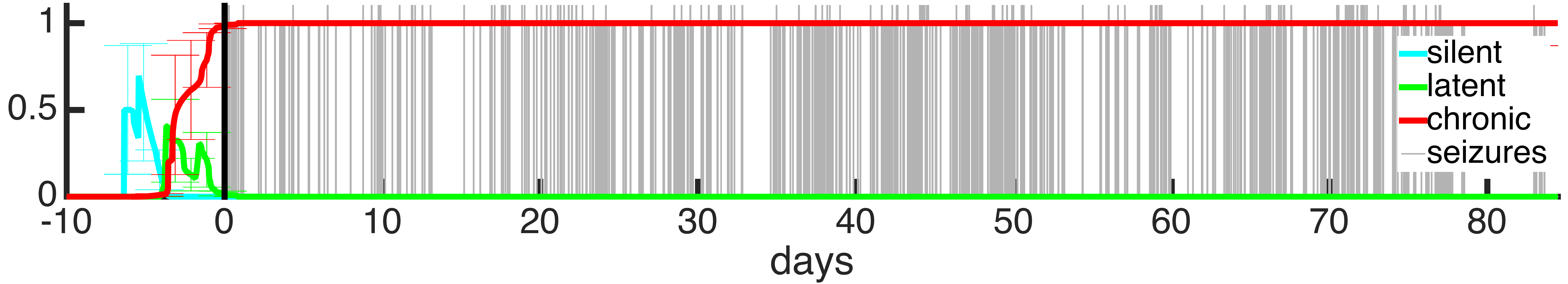}
  }
  \caption{Median probability $\overline{\atk{\pi}}(c)$ of being in condition $c$ for the epileptic
    animals. Day 0 (black bar) corresponds to the first spontaneous seizure across all
    animals. Seizure count (clipped at 1) is displayed in light grey.
    \label{before_srs}}
\end{figure*}

As an alternative, we used two physiologically meaningful temporal landmarks to align the
time series $\left\{\atk{\pi}_r(c), k = 0,1,\ldots\right\}$ of the different animals before
averaging. We used the time of the insult (status epilepticus) to study the transition from baseline
to silent, and we also used the onset of the first spontaneous seizure to study the transition from
latent to chronic. As shown in Figs.~\ref{after_se}-left and Fig.~\ref{before_srs}, this led to two
different median probability profiles for each condition. In each figure, the origin corresponds to
the physiological landmark used to realign the time series.  To help visually compare the two
figures, we added the latent period to Fig.~\ref{after_se}-left, and the silent period to
Fig.~\ref{before_srs}.  We note that the median probability for the silent and latent periods are
different in both figures: indeed, the large variation in the onset of the first spontaneous seizure
leads to different realignment results.

{\noindent \bf Epileptic animals.} The baseline and chronic periods were well defined for the 12 epileptic
animals (see Fig.~\ref{after_se}-left and \ref{before_srs}), and corresponded to the period before
status epilepticus, and the period after the first spontaneous seizure, respectively. The {\em
  sensitivity} was found to be 1 for both the baseline and chronic periods. The {\em specificity}
was 1 for the baseline period, and 0.59 for the chronic period, indicating that the probability
$\atk{\pi}(3)$ rises to 1 before the first spontaneous seizure (see Fig.~\ref{before_srs}).  Indeed,
the biomarker is able to detect changes in the evoked potential $\atk{h}$ that are indicative of neuronal
alterations leading to epilepsy, before the first spontaneous seizure.\\
{\noindent \bf Non-epileptic animals.} Five of the 17 animals, which received the convulsant (pilocarpine),
never developed spontaneous seizures. After the injections of paraldehyde, and the subsequent
recovery from status epilepticus, these animals returned to a baseline condition. These animals were
interesting, as they allowed us to address a common limitation of machine learning based biomarkers
with regard to specificity of the biomarker to the general population. The biomarker was able to
predict the return to baseline condition (see Fig.~\ref{after_se}-right) in spite of the fact that the
training dataset never included non-epileptic rats. This result is significant since it indicates
that the low-dimensional clusters formed by embedding the time-delay wavelet coordinates have 
a universal aspect that can be reliably used to decode the status of the disease.

{\noindent \bf Control animals.} The 5 rats that received no drugs where classified as being in the baseline
condition with probability 1 at all time (results not shown). The 2 rats that received all drug
injections except for pilocarpine, were also classified as being in the baseline conditions at all
time, except for a brief suppression of amplitude, classified as silent (results not shown).

\section*{Acknowledgement}
FGM was supported in part by NSF DMS 1407340.
\bibliographystyle{IEEEtran}
\bibliography{/Users/francois/LaTeX/Bib/biblio}

\begin{thebibliography}{10}
\providecommand{\url}[1]{#1}
\csname url@samestyle\endcsname
\providecommand{\newblock}{\relax}
\providecommand{\bibinfo}[2]{#2}
\providecommand{\BIBentrySTDinterwordspacing}{\spaceskip=0pt\relax}
\providecommand{\BIBentryALTinterwordstretchfactor}{4}
\providecommand{\BIBentryALTinterwordspacing}{\spaceskip=\fontdimen2\font plus
\BIBentryALTinterwordstretchfactor\fontdimen3\font minus
  \fontdimen4\font\relax}
\providecommand{\BIBforeignlanguage}[2]{{%
\expandafter\ifx\csname l@#1\endcsname\relax
\typeout{** WARNING: IEEEtran.bst: No hyphenation pattern has been}%
\typeout{** loaded for the language `#1'. Using the pattern for}%
\typeout{** the default language instead.}%
\else
\language=\csname l@#1\endcsname
\fi
#2}}
\providecommand{\BIBdecl}{\relax}
\BIBdecl

\bibitem{pitkanen11}
A.~Pitk{\"a}nen and K.~Lukasiuk, ``Mechanisms of epileptogenesis and potential
  treatment targets,'' \emph{The Lancet Neurology}, vol.~10, no.~2, pp.
  173--186, 2011.

\bibitem{lukasiuk14}
K.~Lukasiuk and A.~J. Becker, ``Molecular biomarkers of epileptogenesis,''
  \emph{Neurotherapeutics}, vol.~11, no.~2, pp. 319--323, 2014.

\bibitem{shultz14}
S.~R. Shultz, T.~J. O'Brien, M.~Stefanidou, and R.~I. Kuzniecky, ``Neuroimaging
  the epileptogenic process,'' \emph{Neurotherapeutics}, vol.~11, no.~2, pp.
  347--357, 2014.

\bibitem{staba14}
R.~Staba, M.~Stead, and G.~Worrell, ``Electrophysiological biomarkers of
  epilepsy,'' \emph{Neurotherapeutics}, vol.~11, no.~2, pp. 334--346, 2014.

\bibitem{huneau13}
C.~Huneau, P.~Benquet, G.~Dieuset, A.~Biraben, B.~Martin, and F.~Wendling,
  ``Shape features of epileptic spikes are a marker of epileptogenesis in
  mice,'' \emph{Epilepsia}, vol.~54, no.~12, pp. 2219--2227, 2013.

\bibitem{engel12}
J.~Engel and F.~da~Silva, ``High-frequency oscillations--where we are and where
  we need to go,'' \emph{Prog. Neurobiol.}, vol.~98, no.~3, pp. 316--318, 2012.

\bibitem{delaprida15}
L.~M. de~la Prida, R.~J. Staba, and J.~A. Dian, ``Conundrums of high-frequency
  oscillations (80--800 hz) in the epileptic brain,'' \emph{Journal of Clinical
  Neurophysiology}, vol.~32, no.~3, pp. 207--219, 2015.

\bibitem{chauviere09}
L.~Chauvi{\`e}re, N.~Rafrafi, C.~Thinus-Blanc, F.~Bartolomei, M.~Esclapez, and
  C.~Bernard, ``Early deficits in spatial memory and theta rhythm in
  experimental temporal lobe epilepsy,'' \emph{J. Neurosci.}, vol.~29, no.~17,
  pp. 5402--5410, 2009.

\bibitem{hein05}
M.~Hein and J.-Y. Audibert, ``Intrinsic dimensionality estimation of
  submanifolds in \mbox{$R^d$},'' in \emph{Proceedings of the 22nd
  international conference on Machine learning}.\hskip 1em plus 0.5em minus
  0.4em\relax ACM, 2005, pp. 289--296.

\bibitem{bengio06}
M.~Ghil, M.~Allen, M.~Dettinger, K.~Ide, D.~Kondrashov, M.~Mann, A.~W.
  Robertson, A.~Saunders, Y.~Tian, F.~Varadi \emph{et~al.}, \emph{Spectral
  dimensionality reduction}.\hskip 1em plus 0.5em minus 0.4em\relax Springer,
  2006.

\bibitem{tipping99}
M.~Tipping and C.~Bishop, ``Mixtures of probabilistic principal component
  analyzers,'' \emph{Neural computation}, vol.~11, no.~2, pp. 443--482, 1999.

\bibitem{cappe09}
O.~Capp{\'e}, E.~Moulines, and T.~Ryd{\'e}n, \emph{Inference in Hidden
  \mbox{Markov} Models}.\hskip 1em plus 0.5em minus 0.4em\relax Springer, 2009.

\end{thebibliography}
\end{document}